\documentclass[twocolumn, twocolappendix]{aastex631}
\setcitestyle{citesep={,}}
\usepackage{xspace}
\usepackage{mathtools}
\usepackage[encapsulated]{CJK}

\newcommand{\JWST}{JWST\xspace}

\newcommand{\zphot}{\ensuremath{z_{\mathrm{phot}}}\xspace}
\newcommand{\zspec}{\ensuremath{z_{\mathrm{spec}}}\xspace}
\newcommand{\prospb}{Prospector-$\beta$\xspace}
\newcommand{\eazy}{\textsc{eazy}\xspace}
\newcommand{\zqualflag}{\texttt{flag\_zspec\_qual}\xspace}
\newcommand{\specqualflag}{\texttt{flag\_successful\_spectrum}\xspace}
\newcommand{\contamflag}{\texttt{flag\_potential\_local\_background\_issue}\xspace}

\begin{document}

\title{The UNCOVER Survey: First Release of Ultradeep JWST/NIRSpec PRISM spectra\\ 
for $\sim$700 galaxies from $\boldsymbol{z\sim 0.3-13}$ in Abell 2744}

\shorttitle{UNCOVER: First Release of Ultradeep JWST/NIRSpec PRISM spectra in Abell 2744}

\shortauthors{Price et al.}

\author[0000-0002-0108-4176]{Sedona H. Price}
\affiliation{Department of Physics and Astronomy and PITT PACC, University of Pittsburgh, Pittsburgh, PA 15260, USA}

\author[0000-0001-5063-8254]{Rachel Bezanson}
\affiliation{Department of Physics and Astronomy and PITT PACC, University of Pittsburgh, Pittsburgh, PA 15260, USA}

\author[0000-0002-2057-5376]{Ivo Labbe}
\affiliation{Centre for Astrophysics and Supercomputing, Swinburne University of Technology, Melbourne, VIC 3122, Australia}

\author[0000-0001-6278-032X]{Lukas J. Furtak}
\affiliation{Physics Department, Ben-Gurion University of the Negev, P.O. Box 653, Be'er-Sheva 84105, Israel}

\author[0000-0002-2380-9801]{Anna de Graaff}
\affiliation{Max-Planck-Institut f\"ur Astronomie, K\"onigstuhl 17, D-69117, Heidelberg, Germany}

\author[0000-0002-5612-3427]{Jenny E. Greene}
\affiliation{Department of Astrophysical Sciences, Princeton University, 4 Ivy Lane, Princeton, NJ 08544, USA}

\author[0000-0002-5588-9156]{Vasily Kokorev}
\affiliation{Department of Astronomy, The University of Texas at Austin, Austin, TX 78712, USA}

\author[0000-0003-4075-7393]{David J. Setton}\thanks{Brinson Prize Fellow}
\affiliation{Department of Astrophysical Sciences, Princeton University, 4 Ivy Lane, Princeton, NJ 08544, USA}

\author[0000-0002-1714-1905]{Katherine A. Suess}
\altaffiliation{NHFP Hubble Fellow}
\affiliation{Kavli Institute for Particle Astrophysics and Cosmology and Department of Physics, Stanford University, Stanford, CA 94305, USA}

\author[0000-0003-2680-005X]{Gabriel Brammer}
\affiliation{Cosmic Dawn Center (DAWN), Niels Bohr Institute, University of Copenhagen, Jagtvej 128, K{\o}benhavn N, DK-2200, Denmark}

\author[0000-0002-7031-2865]{Sam E. Cutler}
\affiliation{Department of Astronomy, University of Massachusetts, Amherst, MA 01003, USA}

\author[0000-0001-6755-1315]{Joel Leja}
\affiliation{Department of Astronomy \& Astrophysics, The Pennsylvania State University, University Park, PA 16802, USA}
\affiliation{Institute for Computational \& Data Sciences, The Pennsylvania State University, University Park, PA 16802, USA}
\affiliation{Institute for Gravitation and the Cosmos, The Pennsylvania State University, University Park, PA 16802, USA}

\author[0000-0002-9651-5716]{Richard Pan}
\affiliation{Department of Physics and Astronomy, Tufts University, 574 Boston Ave., Medford, MA 02155, USA}

\author[0000-0001-9269-5046]{Bingjie Wang (\begin{CJK*}{UTF8}{gbsn}王冰洁\ignorespacesafterend\end{CJK*})}
\affiliation{Department of Astronomy \& Astrophysics, The Pennsylvania State University, University Park, PA 16802, USA}
\affiliation{Institute for Computational \& Data Sciences, The Pennsylvania State University, University Park, PA 16802, USA}
\affiliation{Institute for Gravitation and the Cosmos, The Pennsylvania State University, University Park, PA 16802, USA}

\author[0000-0003-1614-196X]{John R. Weaver}
\affiliation{Department of Astronomy, University of Massachusetts, Amherst, MA 01003, USA}

\author[0000-0001-7160-3632]{Katherine E. Whitaker}
\affiliation{Department of Astronomy, University of Massachusetts, Amherst, MA 01003, USA}
\affiliation{Cosmic Dawn Center (DAWN), Denmark} 


\author[0000-0002-7570-0824]{Hakim Atek}
\affiliation{Institut d'Astrophysique de Paris, CNRS, Sorbonne Universit\'e, 98bis Boulevard Arago, 75014, Paris, France}

\author[0000-0002-6523-9536]{Adam J.\ Burgasser}
\affiliation{Department of Astronomy \& Astrophysics, UC San Diego, La Jolla, CA, USA}

\author[0009-0009-9795-6167]{Iryna Chemerynska} 
\affiliation{Institut d'Astrophysique de Paris, CNRS, Sorbonne Universit\'e, 98bis Boulevard Arago, 75014, Paris, France}

\author[0000-0001-8460-1564]{Pratika Dayal}
\affiliation{Kapteyn Astronomical Institute, University of Groningen, 9700 AV Groningen, The Netherlands}

\author[0000-0002-1109-1919]{Robert Feldmann}
\affiliation{Department of Astrophysics, University of Zurich, CH-8057, Switzerland}

\author[0000-0003-4264-3381]{Natascha M. {F\"orster Schreiber}}
\affiliation{Max-Planck-Institut f\"ur extraterrestrische Physik, Giessenbachstrasse 1, 85748 Garching, Germany}

\author[0000-0001-7440-8832]{Yoshinobu Fudamoto} 
\affiliation{Center for Frontier Science, Chiba University, 1-33 Yayoi-cho, Inage-ku, Chiba 263-8522, Japan}

\author[0000-0001-7201-5066]{Seiji Fujimoto}\altaffiliation{NHFP Hubble Fellow}
\affiliation{Department of Astronomy, The University of Texas at Austin, Austin, TX 78712, USA}

\author[0000-0002-3254-9044]{Karl Glazebrook}\affiliation{Centre for Astrophysics and Supercomputing, Swinburne University of Technology, PO Box 218, Hawthorn, VIC 3122, Australia}

\author[0000-0003-4700-663X]{Andy D. Goulding}
\affiliation{Department of Astrophysical Sciences, Princeton University, 4 Ivy Lane, Princeton, NJ 08544, USA}

\author[0000-0002-3475-7648]{Gourav Khullar}
\affiliation{Department of Physics and Astronomy and PITT PACC, University of Pittsburgh, Pittsburgh, PA 15260, USA}

\author[0000-0002-7613-9872]{Mariska Kriek}
\affiliation{Leiden Observatory, Leiden University, P.O.Box 9513, NL-2300 AA Leiden, The Netherlands}

\author[0000-0001-9002-3502]{Danilo Marchesini}
\affiliation{Department of Physics and Astronomy, Tufts University, 574 Boston Ave., Medford, MA 02155, USA}

\author[0000-0003-0695-4414]{Michael V. Maseda}\affiliation{Department of Astronomy, University of Wisconsin-Madison, 475 N. Charter St., Madison, WI 53706, USA}

\author[0000-0001-8367-6265]{Tim B. Miller}
\affiliation{Center for Interdisciplinary Exploration and Research in Astrophysics (CIERA), Northwestern University,1800 Sherman Ave, Evanston, IL 60201, USA}

\author[0000-0002-9330-9108]{Adam Muzzin}
\affiliation{Department of Physics and Astronomy, York University, 4700 Keele Street, Toronto, Ontario, ON MJ3 1P3, Canada}

\author[0000-0003-2804-0648 ]{Themiya Nanayakkara}
\affiliation{Centre for Astrophysics and Supercomputing, Swinburne University of Technology, PO Box 218, Hawthorn, VIC 3122, Australia}

\author[0000-0002-7524-374X]{Erica Nelson}
\affiliation{Department for Astrophysical and Planetary Science, University of Colorado, Boulder, CO 80309, USA}

\author[0000-0001-5851-6649]{Pascal A.\ Oesch}
\affiliation{Department of Astronomy, University of Geneva, Chemin Pegasi 51, 1290 Versoix, Switzerland}
\affiliation{Cosmic Dawn Center (DAWN), Niels Bohr Institute, University of Copenhagen, Jagtvej 128, K{\o}benhavn N, DK-2200, Denmark}

\author[0009-0007-1787-2306]{Heath Shipley}
\affiliation{Department of Physics, Texas State University, San Marcos, TX 78666, USA}

\author[0000-0001-8034-7802]{Renske Smit}
\affiliation{Astrophysics Research Institute, Liverpool John Moores University, 146 Brownlow Hill, Liverpool L3 5RF, UK}

\author[0000-0002-5522-9107]{Edward N.\ Taylor}
\affiliation{Centre for Astrophysics and Supercomputing, Swinburne University of Technology, Melbourne, VIC 3122, Australia}

\author[0000-0002-8282-9888]{Pieter van Dokkum}
\affiliation{Astronomy Department, Yale University, 219 Prospect St,
New Haven, CT 06511, USA}

\author[0000-0003-2919-7495]{Christina C.\ Williams}
\affiliation{NSF’s National Optical-Infrared Astronomy Research Laboratory, 950 North Cherry Avenue, Tucson, AZ 85719, USA}

\author[0000-0002-0350-4488]{Adi Zitrin}
\affiliation{Physics Department, Ben-Gurion University of the Negev, P.O. Box 653, Be'er-Sheva 84105, Israel}


\begin{abstract}
We present the design and observations 
of low resolution \JWST/NIRSpec PRISM spectroscopy from the Ultradeep NIRSpec and NIRCam ObserVations before the Epoch of Reionization (UNCOVER) Cycle 1 \JWST Treasury program. 
Targets are selected using \JWST/NIRCam photometry from UNCOVER and other programs, and cover a wide range of categories and redshifts to ensure the legacy value of the survey. 
These categories include 
the first galaxies at $z\gtrsim10$, 
faint galaxies during the Epoch of Reionization ($z\sim6-8$), 
high redshift AGN ($z\gtrsim6$), 
Population III star candidates, 
distant quiescent and dusty galaxies ($1\lesssim{}z\lesssim 6$), 
and filler galaxies sampling redshift--color--magnitude space 
from $z\sim{}0.1-13$. 
Seven NIRSpec MSA masks across the extended Abell 2744 cluster were observed, along 
with NIRCam parallel imaging in 8 filters 
(F090W, F115W, F150W, F200W, F277W, F356W, F410M, F444W, F480M) over a total area of $\sim$26 arcmin$^2$, overlapping existing HST coverage 
from programs including the Hubble Frontier Fields and BUFFALO.  
We successfully observed 553 objects down to $m_{\mathrm{F444W}}\sim30\mathrm{AB}$, and by 
leveraging mask overlaps, we reach total on-target exposure times ranging from $2.4-16.7\mathrm{h}$. 
We demonstrate the success rate and distribution of confirmed redshifts, 
and also highlight the rich information revealed by these ultradeep spectra for a subset of our targets. 
An updated lens model of Abell 2744 is also presented, including 14 additional spectroscopic redshifts and finding a total cluster mass of $M_{\mathrm{SL}}=(2.1\pm0.3)\times10^{15}\,\mathrm{M}_{\odot}$.
We publicly release reduced 1D and 2D spectra for all  objects observed in Summer 2023 along with a  spectroscopic redshift catalog and the updated lens model of the cluster (\url{https://jwst-uncover.github.io/DR4.html}). 
\end{abstract}


\section{Introduction} 
\label{sec:intro}

Deep JWST imaging from early programs has already begun to revolutionize our understanding of the faint, distant universe. The observatory has met or exceeded nearly every pre-flight expectation \citep{Rieke23}, and early data has 
enabled us to find and begin characterizing many galaxy populations that were previously inaccessible: from the first generation of galaxies at Cosmic Dawn 
\citep[e.g.,][]{Naidu22, Atek23, Finkelstein23, Robertson23a, Robertson23b, Casey23}, 
to the faint galaxies driving the reionization of the universe at $z\sim6-9$ \citep[e.g.,][]{Perez-Gonzalez23}, 
to early quiescent galaxies at $z\sim3-5$ \citep[e.g.,][]{Carnall23b, Valentino23}. 
JWST imaging also provides new insights into 
galaxies' detailed structures 
\citep[at $z\lesssim6$; e.g.,][among many others]{Ferreira22, Ferreira23, Kartaltepe23, Martorano23, Nelson23, vanderWel24}, 
including 
reaching low stellar masses approaching those of the dwarf galaxy population 
\citep[$M_*\sim10^6\mathrm{M_{\odot}}$, e.g.,][]{Suess23} 
and revealing 
the structures of heavily dust-obscured galaxies which were previously observable only in the sub-millimeter \citep[e.g.,][]{Kokorev23, Price23, Wu23}. 
Ultradeep JWST imaging has additionally enabled detections of possible globular clusters as early as $z\sim1.4$ \citep[e.g.,][]{Mowla22,Claeyssens23, Forbes23}, 
as well as more detailed studies of globular clusters 
within galaxies out to at least $z\sim0.3$ \citep[e.g.,][]{Harris23, Harris24}. 
Early JWST imaging has also yielded surprises, including larger than anticipated numbers of very luminous early galaxies 
\citep[e.g.,][]{Naidu22,Atek23, Austin23, Bradley23, Finkelstein23,  Adams23, Casey23, Chemerynska24, Robertson23b}
and an unexpected, relatively numerous population of obscured active galactic nuclei (AGN) candidates at high redshift \citep[e.g.,][]{Labbe23_lrd, Furtak23d, 
Barro24, 
Kokorev24, Williams23}.

Taking the next step in exploring these newly uncovered parameter spaces requires leveraging JWST's spectroscopic capabilities to both confirm galaxies' redshifts and to probe their internal physical properties in detail. 
Even with the high sensitivity of JWST/NIRSpec \citep{Boker23}, 
pushing to the most distant and faint regimes is best accomplished with very deep observations in cluster fields, where the strong gravitational lensing boost reaches intrinsically fainter populations by $1-2$ magnitudes relative to blank fields. 
Complementing the aforementioned imaging results, spectra from early \JWST programs have already 
revealed new discoveries and unprecedented measurements. 
Results from this early spectroscopy include confirming the redshifts and properties of galaxies at $z\gtrsim9$ \citep[e.g.,][]{ArrabalHaro23a, ArrabalHaro23b, Curtis-Lake23,RobertsBorsani23}, 
and confirming and characterizing high-redshift obscured AGN \citep[e.g.,][]{Harikane23,Maiolino23,Matthee23} as well as quiescent galaxies at $z\gtrsim3$ 
\citep[e.g.,][]{Carnall23, deGraaff24, Glazebrook24, Carnall24}.

The Ultradeep NIRSpec and NIRCam ObserVations before the Epoch of Reionization (UNCOVER) Cycle 1 Treasury survey 
\citep{Bezanson24}
was designed to collect these deep spectra early in the JWST mission. UNCOVER was designed to obtain 
ultradeep, multiband NIRCam imaging, photometrically detect and characterize galaxies down to $\mathrm{mag_{F444W}}\!\sim30\,$AB \citep{Weaver24}, and then select targets from these newly-observable populations for follow up ultradeep NIRSpec/PRISM multi-object spectroscopy \citep{Ferruit22}. 
The low resolution PRISM mode provides both high sensitivity and wide spectral coverage 
\citep[i.e.,][]{Boker23}, enabling us to constrain continuum breaks down to $\sim29$AB and measure rest-frame ultraviolet (UV) to near infrared (NIR) emission and absorption features 
raging from galaxies within the cluster itself at $z\sim0.3$ out to the earliest epochs at $z\gtrsim10$. 
Early UNCOVER spectroscopic results already address many of these aims, including finding objects among the first generation of galaxies 
\citep[e.g.,][]{Wang23}, characterizing distant obscured AGN \citep[e.g.,][]{Greene24}, and uncovering early quiescent galaxy formation \citep[e.g.,][]{Setton24}.

In this paper we present an overview of the UNCOVER NIRSpec/PRISM spectroscopic observations of 668 targets 
in the Abell 2744 strong lensing cluster field, as well as our coordinated parallel NIRCam imaging which overlaps with existing HST observations from the Hubble Frontier Fields \citep[HFF;][]{Lotz17}  
and BUFFALO \citep{Steinhardt20} programs. 
We detail the target selection and mask design and the observations (Sec.~\ref{sec:obs}), and the spectroscopic reduction and redshift measurements  (Sec.~\ref{sec:reduction}). 
We also present the redshift success rate and distribution of measured redshifts, and discuss example cases of spectra addressing the scientific objectives of the UNCOVER survey (Sec.~\ref{sec:disc}). 
This paper accompanies the public release of early 
reduced NIRSpec/PRISM spectra, spectroscopic redshifts, and 
the NIRCam parallel imaging. 
All magnitudes given are in the AB system \citep{Oke74}.

\section{Spectroscopic Observations}
\label{sec:obs}

\subsection{Target Selection}
\label{sec:targets}

Targets are primarily selected from photometric catalogs constructed from all publicly available HST and JWST imaging over Abell~2744 as of June 2023. 
The JWST/NIRCam observations are: UNCOVER (PIs Labbe \& Bezanson, JWST-GO-2561; \citealt{Bezanson24}), the Early Release Science program GLASS (PI: Treu, JWST-ERS-1324; \citealt{Treu22}), and a Director's Discretionary program (PI: Chen, JWST-DD-2756), providing a total of 8 filters: F090W, F115W, F150W, F200W, F277W, F356W, F410M, and F444W.  The archival HST data consists of HST-GO-11689 (PI: Dupke), HST-GO-13386 (PI: Rodney), HST-DD-13495 (PI: Lotz; \citealt{Lotz17}), and HST-GO-15117 (PI: Steinhardt; \citealt{Steinhardt20}), providing coverage in 7 filters: F435W, F606W, F814W, F105W, F125W, F140W, and F160W. 
The majority of the targets are selected from the UNCOVER NIRCam-selected 
catalog (as presented in \citealt{Weaver24}), 
using internal version v2.2.0. 
This version, 
containing $\sim50,000$ objects down to a combined long-wavelength (LW; F277W+F356W+F444W) 
depth of $\sim30.5\mathrm{AB}$ in the deepest regions, 
included improved treatment of PSF-homogenization and estimates of total magnitudes compared to the initial public DR1 (January 2023).\footnote{The published versions of \citet{Bezanson24} and \citet{Weaver24} include further improvements, corresponding to public data release DR2 (equivalent to internal release v3.0.1).
Photometric redshifts and stellar masses were derived using \eazy (\citealt{Brammer08}; see \citealt{Weaver24}) and \prospb (\citealt{Wang23b}; see \citealt{Wang24}).}
While selecting targets, UNCOVER stellar population modeling including \prospb and \eazy 
were considered (as in \citealt{Weaver24, Wang24}). 
However, the default UNCOVER catalogs excluded a small number of interesting sources, e.g., highly lensed, multiply imaged and/or shredded objects. In these cases, targets were added by hand (with target IDs \mbox{$>60000$}). 
Furthermore, a subset of the targets were selected based on information from other wavelengths, including ALMA sub-mm/mm  (DUALZ, PI: Fujimoto, \citealt{Fujimoto23b}; ALCS, PI: Kohno, \citealt{Fujimoto23c}; ALMA Frontier Fields, PI: Bauer; \citealt{MunozArancibia23}) and Chandra X-ray (e.g., \citealt{Bogdan24}) observations.

For target selection, the updated version of the \citet{Furtak23} analytic lens model of Abell~2744 was used (\texttt{v1.1}).\footnote{The \texttt{v1.1} deflection maps are publicly available on the UNCOVER website:  \url{https://jwst-uncover.github.io/DR2.html\#LensingMaps}.}
This version includes one additional multiple image system in the northern sub-structure (system 82), and more importantly, an additional spectroscopic redshift in the north-western sub-structure from new VLT/MUSE observations of the cluster 
(system 68 at $z=2.584$, \citealt{Bergamini23b}; see also Appendix~\ref{app:multiple-images}). The \texttt{v1.1} lens model achieved a lens plane average image reproduction root-mean-square (RMS) of $\Delta_{\mathrm{RMS}}=0.51\arcsec$.

\begin{figure}[t!]
\centering
\includegraphics[width=0.4725\textwidth]{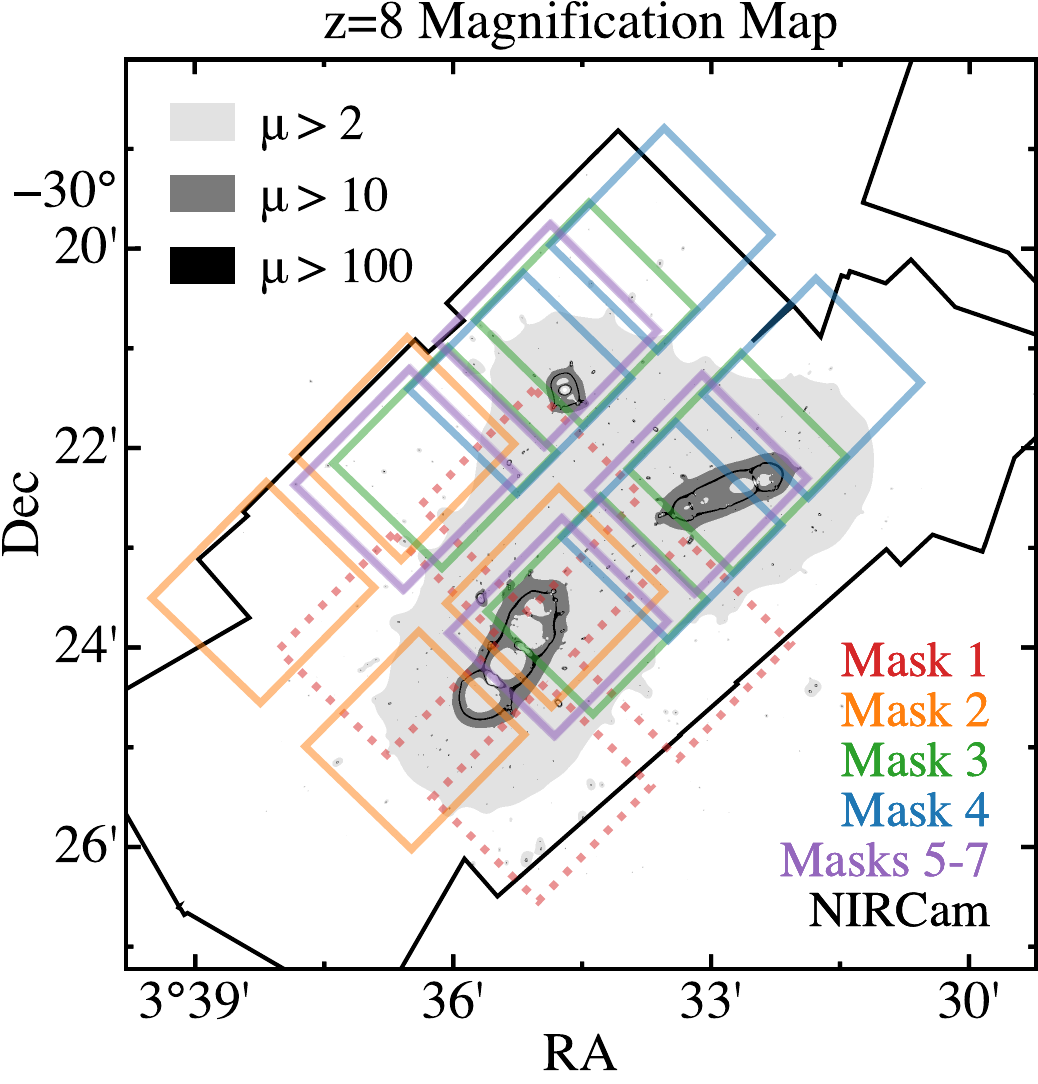}
\vglue -2pt
\caption{
UNCOVER NIRSpec MSA mask footprints within the Abell 2744 cluster field. 
Shaded regions denote the regions of magnification $\mu>2,10,100$ 
(grayscale, light to dark) 
from the updated UNCOVER lensing maps (\texttt{v2.0}) for a source at redshift $z_{\mathrm{s}}=8$, 
and existing NIRCam coverage (from Cycle 1 imaging) is shown with the black outline. 
The masks, shown with colored outlines, span most of the imaging footprint 
over a range of low- and high-magnification regions. 
The electrical short-impacted Mask 1 is marked with a dotted outline. 
(Note Masks 5--7 have near-complete overlap.)
} 
\label{fig:MSA_footprints}
\end{figure}

As the UNCOVER science goals cover a wide range of topics, including potentially risky unknown-unknowns, the final spectroscopic targeting is complex. 
The prioritization scheme for assigning targets to masks is as follows. 
Categories corresponding to the originally proposed science cases (see \citealt{Bezanson24}) are roughly prioritized corresponding to rarity and scientific value: 1) any $z{>}12$ candidates, 2) $z{>}9$ galaxies prioritized by brightness, 3) Pop III candidate sources, 4) faint highly magnified $6{<}z{<}7$ galaxies,\footnote{We emphasize that many of the faint, highly magnified sources remain very small and are compact sources for the NIRSpec microshutter array (and not ``arcs'').} 
5) $z{>}4$ quiescent galaxies, 6) $z{>}6$ AGN, 7) $z{>}4$ dusty galaxies, and other galaxies with ALMA detections (e.g., \citealt{Fujimoto23b}), 8) low mass quiescent galaxies at $1{<}z{<}6$, 9) any unusual or unexpected sources, 10) extreme emission line galaxies, and finally 11) mass-selected ``filler'' galaxies sampled in bins of redshift, mass, and 
F150W$-$LW color (using the LW noise equalized-F277W+F356W+F444W image, and  \eazy-derived mass and redshifts). 
For these filler targets, the numerical priority class $n$ was set to be proportional to the $\log^2$ inverse of the cumulative surface density in each property. As the mask design software eMPT \citep{2023A&A...672A..40B} maps priority class $n$ to weight according to a $1/2^n$ weighting scheme, this approximately equates to an importance sampling scheme that is flat in color, magnitude, and redshift (i.e., sparsely sampling regions of parameter space with many objects, and densely sampling where objects are less common).

\setlength{\tabcolsep}{4pt}
\begin{deluxetable}{cccccc}[t]
\tabletypesize{\footnotesize}
\tablecolumns{6}
\tablewidth{0pt}
\tablecaption{NIRSpec MSA Masks\label{tab:exposures}}
\tablehead{
\colhead{Mask} 
& \colhead{Exposure} 
& \colhead{RA$_{\mathrm{center}}$} 
& \colhead{Dec$_{\mathrm{center}}$} 
& \colhead{PA$_{\mathrm{MSA}}$} 
& \colhead{N$_{\mathrm{target}}$}
\\
\colhead{(1)} 
& \colhead{(2)}
& \multicolumn{2}{c}{(3)} 
& \colhead{(4)} 
& \colhead{(5)} 
}
\startdata
1 & 
\phm{\tablenotemark{\scriptsize{a}}}2.6h\tablenotemark{\scriptsize{a}}
& 3.5839128 & -30.3998611 &  44.5711$^{\circ}$ & 129 \\
2 & 2.6h & 3.6084098 & -30.3911336 & 44.5548$^{\circ}$ &  116 \\
3 & 
\phm{\tablenotemark{\scriptsize{b}}}2.4h\tablenotemark{\scriptsize{b}}
& 3.5732805 & -30.3686750 &  44.5568$^{\circ}$ & 136 \\
4 & 4.4h & 3.5586419 & -30.3564067 & 44.5719$^{\circ}$ &   146 \\
5 & 4.4h & 3.5808445 & -30.3723050 & 44.5608$^{\circ}$ &  144 \\
6 & 4.4h & 3.5803516 & -30.3721636 & 44.5611$^{\circ}$ &  147 \\
7 & 2.9h & 3.5808445 & -30.3723050 & 44.5608$^{\circ}$ &  146 \\
\enddata
\tablecomments{
The sample includes 668 unique targets, with some targets on multiple masks. 
\\
Columns: 
(1) Mask number.  
(2) Mask exposure time (hours). 
(3) Mask center Right Ascension and Declination (J2000).
(4) MSA position angle (deg). 
(5) Number of targets on mask. 
}
\tablenotemark{\scriptsize{a}}: 
The effective total exposure time for Mask 1 is much shorter than the on-sky time, given the electrical short (see Sec.~\ref{sec:design}). Repeat observations of Mask 1 
were taken 30-31 July 2024. \\
\tablenotemark{\scriptsize{b}}: 
The final frame in Visit 3 (Mask 3) for both detectors was lost due to the SSR drive exception. 
\vglue -20pt
\end{deluxetable}
\setlength{\tabcolsep}{2pt}

\subsection{Mask Designs \& Observations}
\label{sec:design}

The NIRSpec/PRISM observations are split into 7 microshutter array (MSA) mask configurations, 
with per-mask exposure times of 
2.6--4.4h (see Table~\ref{tab:exposures}). 
As shown in Figure~\ref{fig:MSA_footprints}, these masks cover the UNCOVER NIRCam primary footprint, 
with overlaps allowing for repeated observations of faint, high-priority targets. 
The masks were designed iteratively using eMPT  \citep{2023A&A...672A..40B}, 
designing each mask in sequence according to target priority, then modifying the priorities to ensure targets requiring deeper integrations are placed on additional masks until the required exposure time is met. This procedure was repeated using hand-specified mask positions until an optimal design (in terms of both number of highest priority targets and total number of targets) was reached.
In total, 668 unique targets are assigned to masks, with total planned exposure times ranging from 2.6 to 17.4 hours.

The NIRSpec observations were taken on 
31 July -- 2 August 2023,  
with a 2-POINT-WITH-NIRCam-SIZE2 dither pattern and a 3 shutter slitlet nod pattern. The NIRSpec NRSIRS2RAPID and NRSIRS2 readout patterns were adopted for Masks 1-3 and 4-7, respectively. 
Coordinated parallel NIRCam imaging was also taken (as described in Appendix~\ref{app:parallel_imaging}). 
The observations were taken with a V3PA angle $\sim$ 266 or NIRSpec MSA aperture PA $\sim44.56$ (see exact values in Table \ref{tab:exposures}), 
to ensure efficient MSA coverage over the UNCOVER NIRCam footprint and 
to overlap the parallel NIRCam imaging with existing HST/ACS and WFC3 observations from the 
HFF \citep{Lotz17} 
and BUFFALO \citep{Steinhardt20} programs.

An electrical short early in Visit 1 severely impacted both detectors, with complete loss for most sources and severely reduced data quality in a minority of objects; 
repeat observations of a slightly modified Mask 1 (due to small differences in PA) were approved, and were observed on 30-31 July 2024. 
Additionally, a solid state recorder (SSR) drive exception (relating to drive space) impacted the Visit 3 observations, leading to a loss of 7\% of the NIRSpec integration time in Mask 3 (1 frame each for both detectors; yielding a total exposure of 2.4h) as well as 66\% of the NIRCam parallel imaging (all of F150W, F200W, F356W, F444W). 
Repeat observations of the NIRCam parallel for Visit 3 
(in all 6 filters, given a probable observing PA change) 
were also approved, and observed on 31 July 2024. 
All repeat observations will be included in a future release.
Given these setbacks, and a small percentage of failed reduction/extractions or other data quality issues, here we present robust spectra for 553 objects, with exposure times of 2.4--16.7 hours.

\section{Spectroscopic Reduction and Redshift Measurements}
\label{sec:reduction}

\subsection{Spectroscopic reduction \& 1D extraction}
\label{sec:redux_details}

The PRISM spectra are reduced using \texttt{msaexp} (v0.8.5; 
\citealt{10.5281/zenodo.7299500}),  \texttt{grizli} (v1.11.9; \citealt{10.5281/zenodo.1146904}), and the JWST \texttt{jwst} pipeline (v1.14.0; \citealt{10.5281/zenodo.10870758}) using the \texttt{jwst\_1241.pmap} reference files. 
Level 1 products are downloaded from MAST\footnote{
Available from: \url{https://dx.doi.org/10.17909/8k5c-xr27}}, and then \texttt{msaexp} (using \texttt{grizli}) runs the \texttt{jwst} stage 1 pipeline, 
inserting the \texttt{snowblind}\footnote{\url{https://github.com/mpi-astronomy/snowblind}} \citep{snowblind_2024} improved ``snowball'' identification and correction procedure after the Jump step. 
\texttt{msaexp} next applies a $1/f$ correction, and, finally, a median pedestal bias offset of the science data (\textsc{sci} extension) and multiplicative scaling factor to the read noise array (\textsc{rnoise} extension) are calculated from empty parts of each exposure that should not have any contribution from sky or source photons. 
Further steps of the \texttt{jwst} stage 2 pipeline are then run to assign the world coordinate system (WCS), flag open microshutters, identify and extract 2D slits, apply slit-level flat-fielding, correct for vignetting of the MSA bars, and apply the photometric calibration.

For this first spectroscopic data release, local background subtraction is performed 
by taking differences of the 2D spectrum arrays at the different telescope nod positions.\footnote{
This scheme can lead to issues for cases with objects/ICL light falling within the flanking shutters, which we discussed later in this section.}
This local background subtraction is performed on the original 2D slitlet cutouts before performing drizzle resampling. 
\texttt{msaexp} then rectifies the 2D spectra from each exposure and resamples them into a final stack with an algorithm analogous to \textsc{drizzle} \citep{fruchter+hook98}, adopting a pixel fraction and wavelength sampling of 1.0. 
In contrast to the STScI \texttt{jwst} drizzle resampling algorithm, the spectra here are only rectified along the columns of the cross-dispersion axis and all wavelength bins are kept fully independent, which eliminates the correlated noise in the dispersion direction that results from a full 2D drizzle resampling.

The final 1D spectra are then extracted from the local background-subtracted 2D spectra using an optimal extraction \citep{Horne86} scheme, modified to account for the variable spatial resolution across the full PRISM wavelength range. 
This modified extraction uses a 2D profile, consisting of independent 1D Gaussians in the cross-dispersion spatial direction over the full wavelength range (and not a single, uniform cross-dispersion profile as used in the standard optimal extraction scheme). 
The 1D spatial cross-dispersion profile at every wavelength is a pixel-integrated Gaussian with width equal to the sum in quadrature of the PSF width (at that wavelength) and an intrinsic cross-dispersion object width. 
This intrinsic width is determined by fitting a model with parameters for both the intrinsic 
profile width and a spatial offset (relative to the position expected from the mask and input catalog metadata)
to the curved traces of the original spectral cutouts (accounting for wavelength-dependent PSF broadening). 
As this fit is performed using the full 2D traces, 
intrinsic widths can be determined 
for objects with detected continua and/or emission lines (even if only emission lines, and not any continuum, is detected). 
For faint objects with little-to-no signal (for either lines or the continuum), the inferred intrinsic width will be close to the initial guess (here, 0.7 pixels).
The final 2D profile(with varying spatial profile as a function of wavelength) used for optimal extraction is rebinned and rectified in the same way as the science data, and the optimally-weighted extraction is performed in the rectified frame. 
This 2D modified optimal extraction profile is included in an extension in the reduced spectra fits files.

Path-loss corrections computed by \texttt{msaexp} are included in the final spectra. 
Using the predicted position of the object within the shutter and assuming the object has an axisymmetric (i.e., round) Gaussian shape with the same intrinsic width measured from the cross-dispersion profiles (as described above), \texttt{msaexp} determines the fraction of light falling within the slit for a source of this size and position relative to a perfectly-centered point source.
This path-loss factor is determined as a function of wavelength (accounting for the variable PSF), 
and is then used to correct the reduced spectra.\footnote{While the path-loss correction does account for wavelength-dependent light losses of extended targets based on their position, the assumptions of axisymmetry (leveraging the cross-dispersion size) and a smooth light distribution will not entirely capture each source's complex morphology. More extensive path-loss modeling based on multiwavelength morphology or scaling to same-aperture-extracted photometry are possible approaches to perform aperture correction that better accounts for these issues.}

We note that we have not attempted to apply 
any aperture corrections (beyond the path-loss correction described above) 
--- that is, any corrections to account for what portion of the target are captured within the slit, particularly with respect to how the targets' photometry has been measured. 
In some cases, it may be beneficial for users to derive a wavelength-dependent aperture correction when jointly modeling photometry and spectroscopy. 
For instance, if a target is very off-center in the dispersion ($x$) direction of the slit or when only a small fraction of the entire object lies within the slit, such a correction would account for differences in photometric and spectroscopic aperture position and size. 
One approach to derive such a correction would be to determine a scaling between the spectrum and photometry (e.g., using a wavelength-dependent polynomial that is fit by comparing the scaled-spectrum photometric-filter-convolved fluxes to the observed fluxes), 
which avoids any detailed calculations (e.g., regarding the slit position or fraction of the object within the slit). 
However, such an approach neglects the presence of color gradients (due to spatially-variable stellar populations or dust, or portions with/without an AGN component, among other possibilities). 
Furthermore, such a spectrum-to-photometry scaling would likely require extrapolation, given the very wide PRISM wavelength coverage.
To facilitate such aperture corrections the user may choose to apply, 
or to enable extracting photometry in the same apertures as the slits, 
we include a catalog detailing the shutter positions of all targets on all masks as part of this data release. 
The source position within the shutter is also stored in the spectrum FITS header, as are the intrinsic cross-dispersion width 
and cross-dispersion extraction center.
\footnote{Specified by ``SRCXPOS'' and ``SRCYPOS'' (in units of shutter width/height, with 0=centered), and ``PROFSIG'' and ``PROFCEN'' (relative to the cross-dispersion center of the 2D spectrum), in units of pixels.}

The ``local'' background subtraction scheme adopted for this release assumes that the flanking shutters of the 3-microshutter slitlets 
represent a reasonable approximation of the 
background within the shutter in which the target is centered. 
This assumption breaks down for large objects which fill multiple adjacent microshutters (resulting in partial ``self-subtraction''), 
or in cases where nearby objects or light from bright cluster galaxies or the ICL fall within these flanking shutters, 
leading to contamination of the background estimate. 
We estimate whether each object is impacted by this issue by determining whether light from any object (either the target itself or a neighbor) or from the ICL/bright cluster galaxies falls within the neighboring shutters (defined as the regions with surface brightness brighter than the target's surface brightness in F444W at its effective radius, from S\'ersic fits if available or from the source extraction half-light radii otherwise). 
A flag \contamflag is included in the released redshift catalog (see Sec.~\ref{sec:zspec} \& Table~\ref{tab:zspec}) to alert users of such potential background-subtraction issues. We emphasize that background-subtraction issues do not preclude the measurement of robust redshifts, nor that the spectrum cannot be used --- this flag is only advisory, indicating that users should examine 
the 2D spectrum and 1D extraction profile to determine if the degree and nature background contamination will impact their analysis. 
Future releases will include global background-subtracted spectra 
(with the background determined accounting for positional variations and any low-order wavelength dependence) 
for all objects, and will be the default recommended spectrum for such large objects and those with other contamination of the neighboring shutters.

Finally, we note that some objects are observed on multiple masks. In the current reduction, all frames of a target are directly combined during the reduction, implicitly assuming that the slitlets of different masks cover the same spatial region of that source. 
We also note that for this first release, spectra from the short-impacted Mask 1 are not reduced. 
The spectra from the repeat observation of Mask 1, and those taken along with the repeat of the Visit 3 NIRCam parallel imaging, will be included in future spectroscopic releases.

\begin{figure}
    \centering
\includegraphics[width=0.5\textwidth]{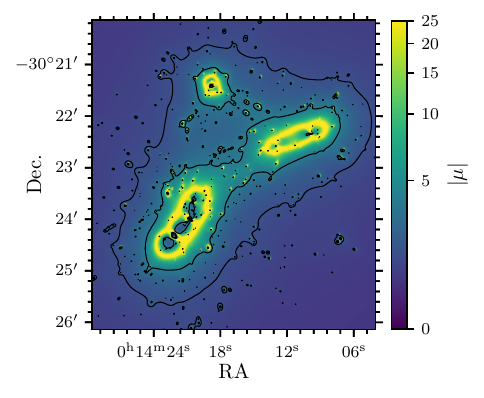}
\vglue -6pt
    \caption{Total magnification map of our new \texttt{v2.0} SL model of Abell 2744 for a source at redshift $z_{\mathrm{s}}=10$. The black contours represent magnification thresholds of $\mu=2$ and $\mu=4$.}
    \label{fig:magnification-map}
\end{figure}

\begin{figure*}[th!]
\centering
\includegraphics[width=0.9\textwidth]{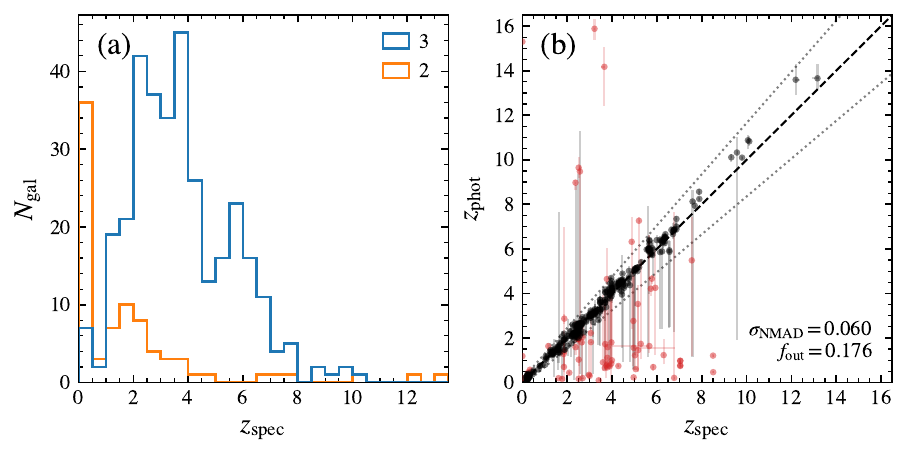}
\vglue -12pt
\caption{
Redshift distribution of spectroscopically confirmed galaxies with 
robust 
redshifts ($\zqualflag\geq2$). 
\textit{Panel a:} Redshift histogram, split by redshift quality flag. 
\textit{Panel b:} 
Spectroscopic versus photometric redshifts, using 
\prospb-derived \zphot from the internal v2.2.0 catalog (used during MSA target selection, including only HST 
and JWST/NIRCam broad-band filters),  with the uncertainties showing the 16, 84$^{\mathrm{th}}$ percentiles. 
Catastrophic outliers (with $|\Delta z| = (\zphot - \zspec)/(1+\zspec) > 0.15$; boundary denoted with dotted lines) are colored red.
} 
\vspace{5pt}
\label{fig:zspec_success_dist}
\end{figure*}

\subsection{Spectroscopic redshifts and line fluxes}
\label{sec:zspec}

The spectroscopic redshifts for this data release are determined from the reduced, full-depth 1D spectra using \texttt{msaexp}. 
First, redshift fits are performed using the \eazy \citep{Brammer08} \texttt{corr\_sfhz\_13} galaxy template set with a wide allowed redshift range 
($z=[0.05,14]$). 
A second redshift fit is then performed with a library of spectral lines and cubic splines for a flexible continuum model, 
restricted within $\pm 0.03(1+z)$ of the template best-fit redshift (or within the range $z=[0.05,14]$ if the template fit failed).  
Models for the emission lines are generated in \texttt{msaexp} as pixel-integrated Gaussians with widths taken from the wavelength-dependent spectral resolution curve provided by STScI and used by the JWST exposure time calculator ($R\sim50$ at 1.5~$\mu\mathrm{m}$, $R\sim300$ at 5~$\mu\mathrm{m}$;\footnote{An upscaling of $\times1.3$ is used to account for the observed PRISM resolution improvement of compact sources compared to the STScI model, as found using \texttt{msafit} \citep{deGraaff24b}.} \href{https://jwst-docs.stsci.edu/jwst-near-infrared-spectrograph/nirspec-instrumentation/nirspec-dispersers-and-filters#NIRSpecDispersersandFilters-DispersioncurvesfortheNIRSpecdispersers}{jwst\_nirspec\_prism\_disp.fits}). 
The prism disperser does not spectrally resolve typical galaxy emission lines, though extremely broad emission (e.g., due to broad-line AGN or outflows) can be resolved.

The spectroscopic redshift for each object is determined as follows: 
(1) from the template fit, for objects with only continuum features based on visual inspection (i.e., only breaks or stellar bumps and no emission lines); or else 
(2) from the lines+splines fit, if at least one emission line is detected with signal-to-noise $\mathrm{S/N}\geq3$ in that fit (and the target was not flagged as only having continuum features in visual inspection); or finally 
(3) from the template fit, from the template fit, if no line is detected. 
The redshift uncertainties for all targets are taken from the 16, 84$^{\mathrm{th}}$ percentiles of the full redshift range template fit (or from the lines+splines fit, if the template fit failed).

The redshift fits are examined by multiple (minimum 3) team members, and flagged based on the number and robustness of the detected spectral features, as described in Table~\ref{tab:zspec}. 
The redshift quality flag, \zqualflag, denotes 
secure redshifts ($=3$; from two or more secure spectral features, e.g., two robustly-detected emission lines, one clear break and one robust emission line, two robustly-detected absorption features), 
solid redshifts ($=2$; from one broad continuum feature, either a break or stellar bump, or from two less robust features, e.g., two marginally-detected emission lines or one marginally detected emission line and a break), 
tentative but unreliable redshifts ($=1$), 
and no redshift solution ($=0$). 
A flag \specqualflag is also included, 
indicating whether the target spectrum was successfully observed and reduced ($=1$) or not ($=0$; due to data quality issues or missing spectra).

In select cases identified during the visual fit inspection (14 objects; 2.5\%), the redshifts are manually refit with alternative settings 
(i.e., multiple robust emission lines where the initial template fits yielded inaccurate redshift estimates; 
noise misidentified as lines when the redshifts are more robustly measured from template fits to continuum breaks) 
or are fixed (the 3 brown dwarfs at $\zspec=0$; see Sec.~\ref{sec:sci_obj}). 
The redshift quality flag is updated based on these modified redshift solutions.

In addition to spectroscopic redshifts, we also determine line fluxes from the  \texttt{msaexp} fits for each object. 
We adopt the values from the same fit as the best-fit redshift 
(described above). The reported line fluxes are not corrected for lensing magnification.

Accompanying this paper, we publicly release reduced spectra and spectroscopic redshifts from the UNCOVER NIRSpec/MSA observations taken in Summer 2023.\footnote{Public release of spectra and redshifts (DR4): \url{https://jwst-uncover.github.io/DR4.html}; also available from Zenodo: \dataset[10.5281/zenodo.13984100]{https://doi.org/10.5281/zenodo.13984100}.}
This data release (UNCOVER DR4) includes the 1D optimally extracted spectra and the 2D spectra with local background subtraction, for all successfully reduced spectra. 
The redshift catalog for this release (described in Table~\ref{tab:zspec} and the downloadable machine readable format version) includes the measured redshifts (if any), redshift and spectra quality flags (including an advisory flag indicating potential issues in the local-background subtraction, as discussed in Sec.~\ref{sec:redux_details} --- here we emphasize again that this flag does not preclude the measurement of robust redshifts, but that it is intended to alert users that they should examine the spectra to determine whether the local background will impact their analysis), 
the total exposure time, and the assigned masks for the full set of targeted objects. 
Subsequent spectroscopic releases will include the 
observations from the repeated Visits 1 \& 3 
and both global- and local-background subtracted spectra 
(optimized for extended and point sources, respectively) 
and an updated redshift catalog.

Reduced mosaics of the NIRCam parallel observations are also available, 
constructed following the same procedures as the cluster NIRCam observations 
(except that modeling and subtraction of bright cluster galaxy and intracluster 
light is not performed). Full details about the parallel mosaics are presented in the 
UNCOVER survey paper \citep{Bezanson24}.

\subsection{Updating the UNCOVER lens model} 
\label{app:SL-model}

We also use the UNCOVER spectroscopy to update the lens model of Abell~2744 presented in \citet{Furtak23} and include the new \texttt{v2.0} model in the data release. As described in detail in Appendix~\ref{app:lens-model}, this model incorporates the UNCOVER DR4 spectroscopic redshifts of multiple images, and all currently available \JWST imaging for cluster member selection. In total, we added 14 spectroscopic redshifts compared to our initial \texttt{v1.0} model. The model is constructed with an updated version of the analytic lens modeling method by \citet{Zitrin15}. 
We refer the reader to Appendix~\ref{app:lens-model} and \citet{Furtak23} for details of the parameterization for our lensing model of Abell~2744. 

With these constraints, the model achieves an average image reproduction error of $\Delta_{\mathrm{RMS}}=0.60\arcsec$, which is slightly better than our \texttt{v1.0} model \citep[$\Delta_{\mathrm{RMS}}=0.66\arcsec$][]{Furtak23}. The critical lines and multiple image positions are shown in Figure~\ref{fig:crit-lines} in Appendix~\ref{app:lens-model} and we show an updated magnification map at source redshift $z_{\mathrm{s}}=10$ in Figure~\ref{fig:magnification-map}. We find the cluster to have a total critical area of $A_{\mathrm{crit}}=0.63\,\mathrm{arcmin}^2$ for a source at $z_{\mathrm{s}}=2$. This translates to an effective Einstein radius of $\theta_{\mathrm{E}}=26.9\arcsec\pm2.7\arcsec$ enclosing a mass of $M(<\theta_{\mathrm{E}})=(1.0\pm0.2)\times10^{14}\,\mathrm{M}_{\odot}$. These also agree well with our measurements from our \texttt{v1.0} model \citep{Furtak23}. Summing the surface mass density over the entire field (see Figure~\ref{fig:crit-lines}), we obtain a total cluster mass of $M_{\mathrm{SL}}=(2.1\pm0.3)\times10^{15}\,\mathrm{M}_{\odot}$. This is comparable to an $M_{200}$ mass and thus places Abell~2744 well within the mass range of typical clusters with the same Einstein radius \citep[e.g.][]{Fox22}.

\begin{figure*}
\centering
\hglue -10pt
\includegraphics[width=0.99\textwidth]{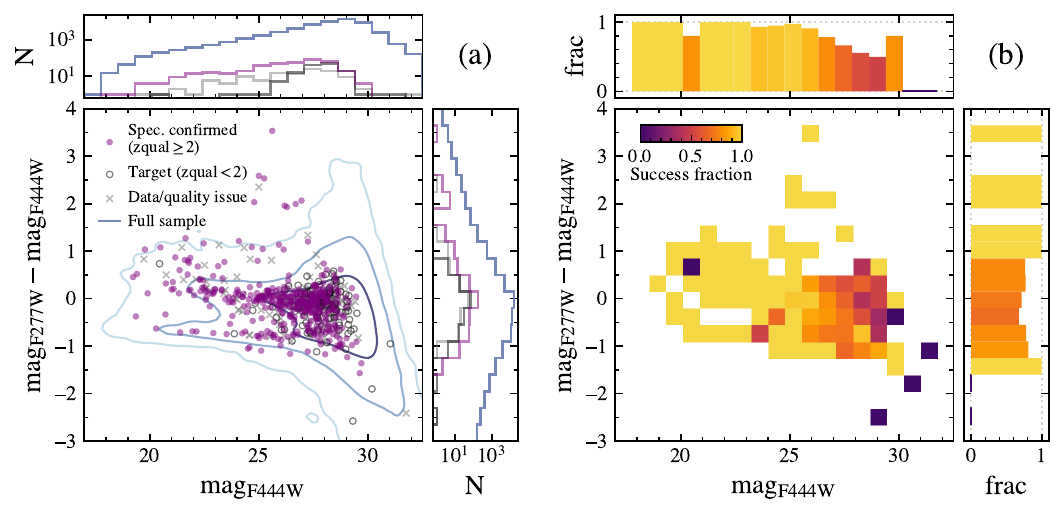}
\vglue -10pt
\caption{Distribution of the spectroscopic sample relative to the full UNCOVER photometric 
catalog (\textit{left}) and 
the redshift measurement success rate (\textit{right}) 
over total F444W magnitude versus F277W--F444W color. 
All values are taken from the internal v2.2.0 catalog (used for designing masks). 
\textit{Panel a:} 
Points indicate the spectroscopically-confirmed objects (\zqualflag$\geq2$; filled purple circles), targets without robust redshifts 
(\zqualflag$<2$; dark gray open circles), and targets with data quality issues (e.g., those on MSA1; gray crosses). 
Contours denote the parent photometric sample distribution 
(with $\texttt{use\_phot}=1$; see \citealt{Weaver24}; 1,2,3$\sigma$ levels). 
Side panels show histograms over $\mathrm{mag_{F444W}}$ and F277W--F444W (line colors the same as points in the main panel). 
Though 
the sample selection incorporates multiple disparate categories, 
overall the targets follow the distribution of the parent sample 
down to $\mathrm{mag_{F444W}}\sim29\,\mathrm{AB}$. 
Successfully-observed targets without measured \zspec 
do not have systematically redder/bluer colors compared to the spectroscopically-confirmed ones. 
The unconfirmed targets are fainter on average than the confirmed objects, though their distribution does overlap 
down to the very faintest magnitudes ($\mathrm{mag_{F444W}}\gtrsim30\,\mathrm{AB}$) 
\textit{Panel b:} 
2D and 1D histograms of the redshift measurement success fraction over $\mathrm{mag_{F444W}}$ and F277W--F444W, 
defined as the fraction of 
objects with robust redshifts over the total number of 
successfully-observed targets. 
The success fraction is very high over most of this space, 
though drops to $\sim30-50\%$ at $\mathrm{mag_{F444W}}\sim29\,\mathrm{AB}$. } 
\vspace{5pt}
\label{fig:sample_color_mag}
\end{figure*}

The \texttt{v2.0} lens model is included in the UNCOVER DR4. The public lensing products include deflection $\alpha$, convergence $\kappa$, shear $\gamma$, magnification $\mu$ and potential $\psi$ maps, normalized to $D_{\mathrm{ds}}/D_{\mathrm{s}}=1$, as well as catalogs of the cluster member galaxies and multiple images used. The JWST cluster member selection and spectroscopic redshifts of multiple images are further detailed in Appendices \ref{app:cluster-members} and \ref{app:multiple-images} respectively. We also updated the UNCOVER photometric and spectroscopic catalogs with magnification and shear parameters from the \texttt{v2.0} model. Individual models of each of the three sub-structures separately are also available on request, each achieving local image reproduction errors of $\Delta_{\mathrm{RMS}}\simeq0.2\arcsec$.

\section{Discussion}
\label{sec:disc}

\subsection{Success rate and redshift distribution for spectroscopically-confirmed objects}
\label{sec:z_dist}

The UNCOVER spectroscopic redshift catalog includes a 74\% success rate, with robust redshifts 
(i.e., defined as $\zqualflag\geq2$; see Table~\ref{tab:zspec} and Sec.~\ref{sec:zspec}) 
for 409 of the 553 targets 
with successfully observed and reduced spectra.
A histogram of the redshift distribution of spectroscopically confirmed targets, split by \zqualflag, is shown 
in Figure~\ref{fig:zspec_success_dist}a. 
We measure secure redshifts 
(based on two or more secure spectral features; $\zqualflag=3$) 
for 327 objects, spanning from $z\sim0.3$ to $z\sim10$. 
The 82 galaxies with solid redshifts 
(based on one broad continuum feature or 2 less robust features; $\zqualflag=2$) 
also span a wide redshift range 
($z\sim0.2-13$). 
This latter category includes most of the targeted galaxies in the Abell 2744 cluster itself, 
as most have very red spectra with no emission lines 
and only a broad stellar bump (resulting in lower redshift precision). 

We compare the spectroscopic and photometric redshifts for our sample of 
spectroscopically-confirmed galaxies in Figure~\ref{fig:zspec_success_dist}b. 
We find the majority of the \prospb-derived \zphot (from the internal v2.2.0 catalog, 
the most up-to-date catalog used during MSA design in early Summer 2023) 
are in good agreement 
with the measured \zspec, with a low normalized median absolute deviation 
$\sigma_{\mathrm{NMAD}}=0.060$. 
However, there is a relatively high fraction of catastrophic photometric redshift outliers 
(with $|\Delta z| = (\zphot - \zspec)/(1+\zspec) > 0.15$; 17.6\%, red circles).

Figure~\ref{fig:sample_color_mag}a shows that 
successfully-observed spectroscopic targets with and without robust redshifts ($\zqualflag\geq \,\mathrm{and}\, <2$, purple filled and gray open circles, respectively)  
have similar F277W--F444W colors. 
On average, targets without robust redshifts are fainter in F444W than spectroscopically-confirmed objects, 
though both have overlapping distributions down to the very faintest magnitudes ($\mathrm{mag_{F444W}}\gtrsim30\,\mathrm{AB}$). 
This is quantified in Figure~\ref{fig:sample_color_mag}b, as the spectroscopic success fraction is very high for bright targets, but drops only to $\sim30-50\%$ at $\mathrm{mag_{F444W}}\sim29\,\mathrm{AB}$ (excepting a few extremely faint targets at $\mathrm{mag_{F444W}}\gtrsim30\,\mathrm{AB}$). 
The distributions of the targets with and without robustly-measured redshifts suggest that 
while low S/N does contribute to failed spectroscopic confirmations, low S/N is not entirely responsible for the failed spectroscopic confirmations. 
Color likewise appears to not drive failed \zspec measurements.

\begin{figure*}[ht!]
\centering
\includegraphics[width=0.95\textwidth]{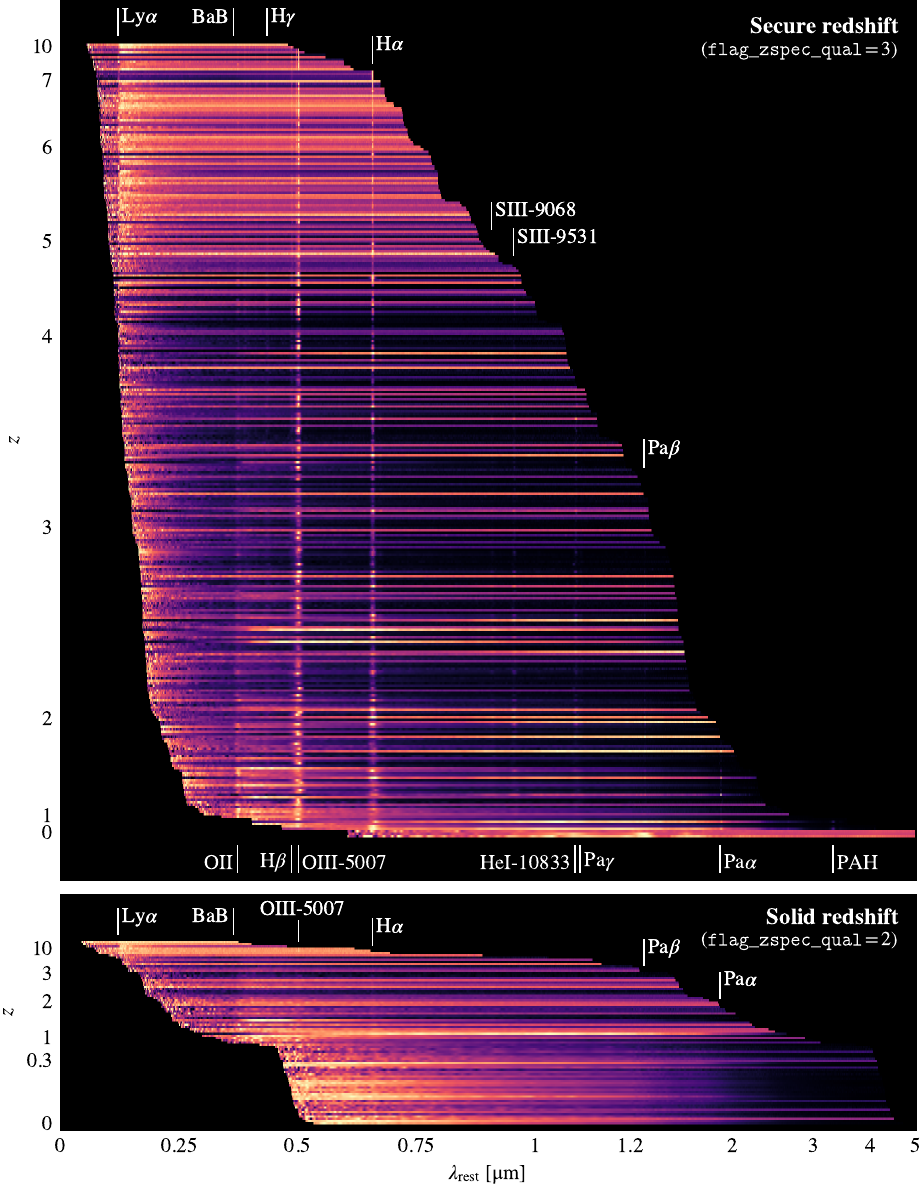}
\caption{
NIRSpec/PRISM spectra for all UNCOVER targets with 
secure ($\mathtt{flag\_zspec\_qual}=3$, $N=327$; \textit{top}) and 
solid ($\mathtt{flag\_zspec\_qual}=2$, $N=82$; \textit{bottom}) redshifts, 
shifted to the restframe and ordered by increasing redshift. 
The wavelength axis is split, with linear and log scaling below and above $1.2\mathrm{\mu{}m}$, respectively. The locations of notable emission and absorption/break features are annotated above and below the spectra.
} 
\label{fig:rf_spectra_z}
\end{figure*}

We similarly find catastrophic photometric redshift failures 
within of the spectroscopically-confirmed sample 
($|\Delta{}z|>0.15$; shown with red points in Figure~\ref{fig:zspec_success_dist}) 
are not primarily driven by low S/N or color, 
as these objects 
exhibit a wide range of magnitudes and F277W--F444W colors 
similar to the complete spectroscopically confirmed sample. 
Preliminary visual inspection suggests 
some of these outliers are due to confusion of the Lyman and Balmer breaks in the \prospb redshift fits; 
the outlier fraction is lower in the region with deep HST/ACS coverage, as rest-frame UV coverage at high redshifts helps to mitigate break confusion. 
Other outliers may be explained by emission line boosting of the broad-band photometry. 
We note that for the few very red targets 
($\mathrm{F277W-F444W\gtrsim1}$), nearly all have catastrophic photometric redshift failures, suggesting additional spectral templates for photometric redshift fitting may be needed to capture the extreme colors of these objects.

A more detailed discussion of photometric redshift outliers relative to the measured \zspec 
will be presented in a forthcoming paper (see also \citealt{Suess24}). 
This will include quantification of photometric redshift improvements by incorporating recently-obtained medium band imaging (from JWST-GO-4111; \citealt{Suess24}) --- which will better account for emission line boosting --- and the inclusion of bluer coverage from NIRCam/F070W.  Adding future deep UV HST/WFC3/UVIS F336W imaging from an approved program (HST-GO-17730; PIs: Whitaker, Bezanson, Leja) will also help to improve photometric redshifts by mitigating break confusion.

\begin{figure*}[th!]
\centering
\includegraphics[width=0.95\textwidth]{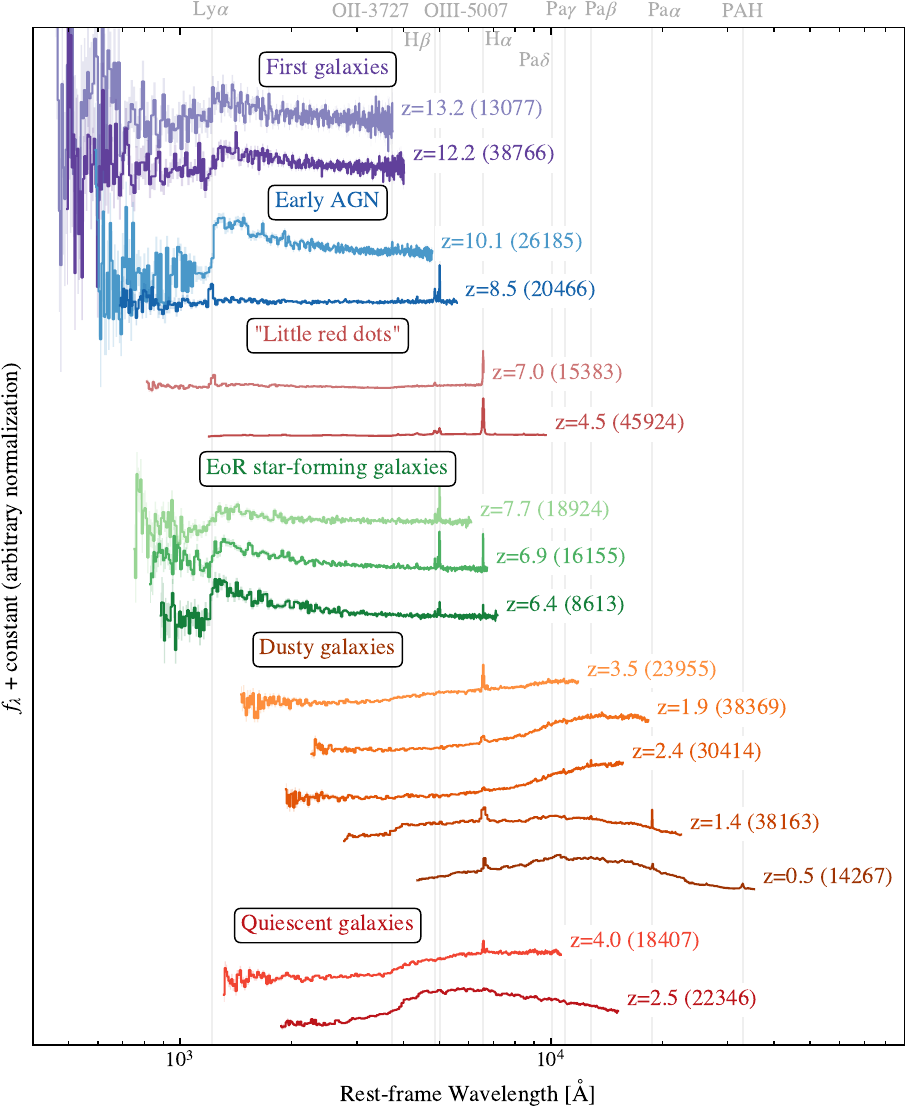}
\caption{
Overview of 1D spectra for a subset of our sample, highlighting key science themes 
addressed by the UNCOVER survey and mask design strategy. 
All spectra are shown in the restframe in $f_{\lambda}$ units (with arbitrary normalization and shifting), with the shaded contour denoting the uncertainty. The redshift and MSA ID of each object are annotated next to the spectra. 
Vertical lines mark the wavelengths of selected emission features. 
} 
\label{fig:spec_gallery}
\end{figure*}

\subsection{Scientific objectives addressed by the UNCOVER spectroscopic sample}
\label{sec:sci_obj}

With a high redshift success rate (74\% of targets with robust redshifts) 
and very deep spectra (up to 16.7h for five individual targets, and up to 38h with multiply lensed images of one object), 
these NIRSpec observations provide a treasure trove for 
studies ranging from galaxies within the Abell 2744 cluster itself out to the first galaxies at Cosmic Dawn. 
We demonstrate the wide range of spectral features seen in the full UNCOVER sample of 
409 galaxies with robust \zspec (\zqualflag$\geq2$)
ranging from $z\sim0.3$ to $z\sim13$ in Figure~\ref{fig:rf_spectra_z}. 
An incredible diversity of features can be seen in these ultradeep spectra: 
Paschen lines, HeI$-10833\mathrm{\AA}$, and the polycyclic aromatic hydrocarbon (PAH) 3.3$\mu$m feature 
are seen in galaxies at the lowest redshifts, and [SIII]$-9068,9531\mathrm{\AA\AA}$ are seen out to $z\sim5$. H$\alpha$ is detected out to $z\sim7$, and H$\beta$ and [OIII]$-4959,5007\mathrm{\AA\AA}$ are seen in all galaxies except those at the very highest redshifts. 
The Balmer break is seen in galaxies at $z\gtrsim1$, while the Lyman break (and Ly$\alpha$) are seen at $z\gtrsim4.5$.

An overview of the science cases addressed by the UNCOVER spectra are highlighted 
in Figure~\ref{fig:spec_gallery}, showing a subset of objects from our sample. 
We have spectroscopically confirmed and begun characterizing 
the rest-frame UV spectra of 10 early galaxies at $z\geq 8.5$ 
(see e.g., \citealt{Fujimoto23}), 
including two among the first generation of galaxies at $\zspec=13.03$ and $\zspec=12.39$ (as presented in \citealt{Wang23}). 
Our sample also features a number of early AGN at $z>6$, including an X-ray luminous AGN at $\zspec=10.1$ (\citealt{Bogdan24}, \citealt{Goulding23}) and a broad-line AGN at $\zspec=8.5$ (\citealt{Kokorev23}). Other targets include a number of dust-reddened, high-redshift objects described as ``little red dots'' (e.g., \citealt{Labbe23_lrd}, \citealt{Furtak23b}, \citealt{Greene24}). 
The PRISM spectra also yield the first spectroscopic constraints on low-mass, low-luminosity galaxies during the Epoch of Reionization ($z\sim6-8$), including direct constraints on the ionizing photon production efficiency that yield evidence that these faint galaxies are the primary drivers of the reionization of the Universe (\citealt{Atek24, Dayal24}), and extending the mass-metallicity relation to the low-mass end \citep{Chemerynska24b}.

Our spectroscopic sample additionally includes a range of dusty galaxies out 
to $z\sim4$, both with (e.g., some of the 
objects presented in \citealt{Kokorev23} and \citealt{Price23}) 
and without ALMA continuum detections 
\citep[from e.g., DUALZ,][]{Fujimoto23b}. 
Two targeted galaxies at low redshift ($z\lesssim0.5$) 
reveal detections of the 3.3$\mu$m PAH emission feature and 
ice absorption features. 
Also targeted are a number of quiescent galaxies extending from low redshift to $z\gtrsim3$. 
This includes a massive, dusty quiescent galaxy confirmed at $\zspec=3.97$ (\citealt{Setton24}), 
with the deep PRISM spectra revealing its detailed star formation history 
that indicates the early formation of its dense stellar core. 
Finally, we obtained spectra for three brown dwarfs located within 
our own Milky Way (\citealt{Langeroodi23}, \citealt{Burgasser24}): 
one explicitly targeted, and two that were selected based on the photometric criteria for dust-reddened ``little red dots'' and AGN at high redshift. These deep spectra reveal the spectral classifications, temperatures, and metallicities, as well as characterizing molecular features within the brown dwarf atmospheres. 


\setlength{\tabcolsep}{4pt}
\begin{splitdeluxetable*}{ccccccccccBccccccccccc}
\label{tab:zspec}
\tabletypesize{\footnotesize}
\tablecolumns{10}
\tablewidth{0pt}
\tablecaption{Redshift catalog from UNCOVER NIRSpec/PRISM spectra}
\tablehead{
\colhead{\texttt{id\_msa}} 
& \colhead{\texttt{ra}} 
& \colhead{\texttt{dec}} 
& \colhead{\texttt{z\_spec}}
& \colhead{\texttt{z\_spec16}}
& \colhead{\texttt{z\_spec50}}
& \colhead{\texttt{z\_spec84}}
& \colhead{\texttt{flag\_}}
& \colhead{\texttt{flag\_}}
& \colhead{\texttt{flag\_}}
& \colhead{\texttt{flag\_}}
& \colhead{\texttt{flag\_}}
& \colhead{\texttt{flag\_break\_}}
& \colhead{\texttt{flag\_}}
& \colhead{\texttt{flag\_}}
& \colhead{\texttt{method\_}}
& \colhead{\texttt{method\_}}
& \colhead{\texttt{texp\_tot}}
& \colhead{\texttt{masks}}
& \colhead{\texttt{id\_DR3}} 
& \colhead{\texttt{sep\_DR3}}
\vspace{-10pt}\\
\colhead{} 
& \colhead{} 
& \colhead{} 
& \colhead{} 
& \colhead{} 
& \colhead{} 
& \colhead{} 
& \colhead{\texttt{zspec\_}} 
& \colhead{\texttt{successful\_}} 
& \colhead{\texttt{potential\_local\_}}
& \colhead{\texttt{emission\_}}
& \colhead{\texttt{line\_and\_}}
& \colhead{\texttt{strong\_abs\_}}
& \colhead{\texttt{break\_}}
& \colhead{\texttt{stellar\_}}
& \colhead{\texttt{best\_}}
& \colhead{\texttt{zuncert}}
& \colhead{}
& \colhead{}
& \colhead{}
& \colhead{}
\vspace{-10pt}\\
\colhead{} 
& \colhead{} 
& \colhead{} 
& \colhead{} 
& \colhead{} 
& \colhead{} 
& \colhead{} 
& \colhead{\texttt{qual}} 
& \colhead{\texttt{spectrum}}
& \colhead{\texttt{background\_issue}}
& \colhead{\texttt{lines}}
& \colhead{\texttt{break}}
& \colhead{\texttt{features\_only}}
& \colhead{\texttt{only}}
& \colhead{\texttt{bump\_only}}
& \colhead{\texttt{zfit}}
& \colhead{}
& \colhead{}
& \colhead{}
& \colhead{}
& \colhead{}
\vspace{-5pt}
\\
\colhead{(1)} 
& \multicolumn{2}{c}{(2)} 
& \colhead{(3)} 
& \colhead{(4)} 
& \colhead{(5)} 
& \colhead{(6)} 
& \colhead{(7)} 
& \colhead{(8)} 
& \colhead{(9)} 
& \colhead{(10)} 
& \colhead{(11)} 
& \colhead{(12)} 
& \colhead{(13)} 
& \colhead{(14)} 
& \colhead{(15)} 
& \colhead{(16)} 
& \colhead{(17)} 
& \colhead{(18)} 
& \colhead{(19)} 
& \colhead{(20)} 
}
\startdata 
 2008 & 3.59242259 & -30.43282858 & --- & --- & --- & --- & 0 & 0 & 1 & --- & --- & --- & --- & --- & --- & --- & 0.0 & 1          & 10065 & 0.009 \\
 2044 & 3.58615595 & -30.43309276 & --- & --- & --- & --- & 0 & 0 & 1 & --- & --- & --- & --- & --- & --- & --- & 0.0 & 1          & 10155 & 0.017 \\
 2354 & 3.58522511 & -30.43136947 & --- & --- & --- & --- & 0 & 0 & 0 & --- & --- & --- & --- & --- & --- & --- & 0.0 & 1          & 10730 & 0.009 \\
 2385 & 3.58064554 & -30.43128230 & --- & --- & --- & --- & 0 & 0 & 1 & --- & --- & --- & --- & --- & --- & --- & 0.0 & 1          & 10787 & 0.021 \\
\ldots & \ldots & \ldots & \ldots & \ldots & \ldots & \ldots & \ldots & \ldots & \ldots & \ldots & \ldots & \ldots & \ldots & \ldots & \ldots & \ldots & \ldots & \ldots & \ldots & \ldots \\
43197 & 3.59312594 & -30.34885302 & 3.791 & 3.788 & 3.793 & 3.798 & 3 & 1 & 0 & 1 & 0 & 0 & 0 & 0 & spl+lines        & templ            & 4.4 & 4          & 55712 & 0.005 \\
43239 & 3.57603883 & -30.34966523 & 0.172 & 0.170 & 0.171 & 0.173 & 2 & 1 & 1 & 0 & 0 & 0 & 0 & 1 & templ            & templ            & 7.3 & 5,7        & 56010 & 0.099 \\
43311 & 3.56381835 & -30.34873405 & 3.291 & 3.285 & 3.289 & 3.294 & 3 & 1 & 0 & 1 & 0 & 0 & 0 & 0 & spl+lines        & templ            & 7.3 & 5,7        & 55788 & 0.002 \\
43388 & 3.56605332 & -30.34854295 & 3.801 & 3.795 & 3.801 & 3.806 & 3 & 1 & 0 & 1 & 0 & 0 & 0 & 0 & spl+lines        & templ            & 7.3 & 5,7        & 55908 & 0.006 \\
\ldots & \ldots & \ldots & \ldots & \ldots & \ldots & \ldots & \ldots & \ldots & \ldots & \ldots & \ldots & \ldots & \ldots & \ldots & \ldots & \ldots & \ldots & \ldots & \ldots & \ldots \\
\enddata
\tablecomments{The full table is available in machine readable format from  \url{https://jwst-uncover.github.io/DR4.html} and \dataset[https://doi.org/10.5281/zenodo.13984100]{https://doi.org/10.5281/zenodo.13984100}. \\
Columns: 
\\
(1) MSA ID (corresponding to internal v2.2.0). 
\\
(2)  Targeted Right Ascension and Declination (internal v2.2.0 catalog; J2000, decimal degrees).
\\
(3) Spectroscopic redshift. 
\\
(4)--(6) 16/50/84th percentile from redshift fit $p(z)$  distribution. 
\\
(7) Redshift quality flag: 
3 = secure, based on two or more secure spectral features 
(e.g., two robustly-detected emission lines, one clear break and one robust emission line, two robustly-detected absorption features); 
2 = solid, based on one broad continuum feature or two less robust features 
(e.g., a break or stellar bump, or two marginally-detected emission lines, or one marginally-detected emission lines and a break); 
1 = tentative but unreliable redshift; 
0 = no redshift. 
\textbf{For analysis, using redshifts with quality flag $\mathbf{=3}$ or $\mathbf{=2}$ is recommended.}
\\
(8) Spectrum flag: 
1 = successfully observed and reduced spectrum;  
0 = no spectrum/data quality issue. 
\\
(9) Local background subtraction issue flag: 1 = objects with 
potential issues from galaxy/ICL light in the neighboring shutters; 0 = no local background subtraction issues. 
This flag does not preclude the measurement of robust redshifts, nor the use of this spectrum for science analysis --- rather this is an advisory flag to alert users to inspect the 2D spectrum and evaluate if this background issue could impact their planned analysis.
\\
(10) Feature flag, for spectra containing two or more emission lines (1=yes, 0=no). 
\\
(11) Feature flag, for spectra containing a break + an emission line (1=yes, 0=no). 
\\
(12) Feature flag, for spectra containing only a break and strong absorption features (1=yes, 0=no). 
\\
(13) Feature flag, for spectra containing only a break (1=yes, 0=no). 
\\
(14) Feature flag, for spectra containing only a stellar bump (1=yes, 0=no). 
\\
(15) Fit method for best-fit redshift. 
\\
(16) Fit method for redshift uncertainties/percentiles. 
\\
(17) Total exposure time (hours). 
\\
(18) List of masks on which each object was included (comma separated string). 
\\
(19) Closest match DR3 ID (\citealt{Suess24}, \citealt{Weaver24}, \citealt{Wang24}).
\\
(20) Separation of DR3 \& MSA RA/Dec, in arcsec. 
}
\end{splitdeluxetable*}
\setlength{\tabcolsep}{2pt}

\section{Final remarks}
\label{sec:finalremarks}
The ultradeep PRISM spectra from the UNCOVER program add immense value to the already rich --- and still growing --- treasure trove of public observations in the Abell 2744 lensing cluster field. 
This first data release of the 1D and 2D spectra (with local background subtraction), along with derived catalogs with quality flags, is publicly available on the survey website (\url{https://jwst-uncover.github.io/DR4.html}). 
Future spectroscopic releases will include 
the repeat observations of MSA1 
and spectra accompanying the Visit 3 repeat of the NIRCam parallel imaging (observed 30-31 July 2024). 
Additional improvements in the reduction and released products will include global background subtraction and more sophisticated modeling of emission lines.

\section*{}
\noindent
We thank the referee for a constructive and insightful report which has improved this manuscript.
This work is based in part on observations made with the NASA/ESA/CSA \emph{James Webb Space Telescope}. The data were obtained from the Mikulski Archive for Space Telescopes at the Space Telescope Science Institute, which is operated by the Association of Universities for Research in Astronomy, Inc., under NASA contract NAS 5-03127 for \JWST. These observations are associated with JWST-GO-2561. Support for program JWST-GO-2561 was provided by NASA through a grant from the Space Telescope Science Institute, which is operated by the Associations of Universities for Research in Astronomy, Incorporated, under NASA contract NAS5-26555. 
The specific observations analyzed can be accessed via \dataset[10.17909/8k5c-xr27]{http://dx.doi.org/10.17909/8k5c-xr27}.
Cloud-based data processing and file storage for this work is provided by the AWS Cloud Credits for Research program. 
The Cosmic Dawn Center is funded by the Danish National Research Foundation (DNRF) under grant \#140.
The BGU lensing group acknowledges support by grant No.~2020750 from the United States-Israel Binational Science Foundation (BSF) and grant No.~2109066 from the United States National Science Foundation (NSF), by the Israel Science Foundation Grant No.~864/23, and by the Ministry of Science \& Technology, Israel.

\facilities{JWST(NIRSpec, NIRCam)}

\software{
    astropy \citep{2013A&A...558A..33A,2018AJ....156..123A,2022ApJ...935..167A}, 
    eMPT \citep{2023A&A...672A..40B}, 
    \texttt{jwst} pipeline (v1.14.0; \citealt{10.5281/zenodo.10870758}), 
    \texttt{msaexp} (v0.8.5; \citealt{10.5281/zenodo.7299500}), 
    \texttt{grizli} (v1.11.9; \citealt{10.5281/zenodo.1146904}), 
    \eazy \citep{Brammer08}, 
    matplotlib \citep{10.1109/MCSE.2007.55},  
    numpy \citep{10.1038/s41586-020-2649-2}, 
    scipy \citep{10.1038/s41592-019-0686-2}, 
    seaborn \citep{waskom17},
    \texttt{snowblind} \citep{snowblind_2024} 
}

\appendix

\section{Parallel NIRCam Imaging}
\label{app:parallel_imaging}

\begin{figure}[t!]
\centering
\includegraphics[width=0.4725\textwidth]{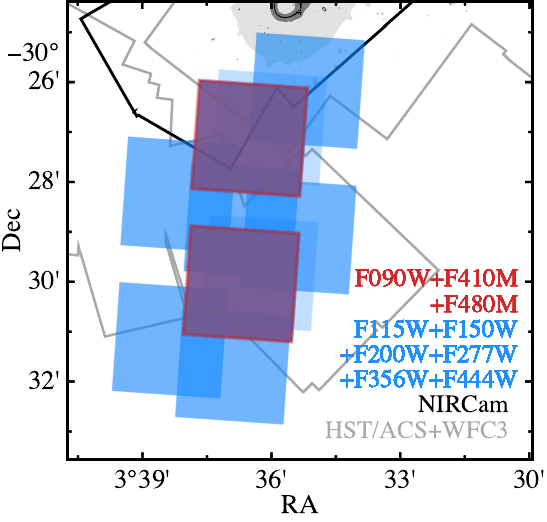}
\vglue -2pt
\caption{
The NIRCam parallel footprints, 
plotted over the existing NIRCam and HST/ACS+WFC3 coverage footprints and the lensing contours as shown in Figure~\ref{fig:MSA_footprints}. 
Coverage of F090W, F410M, and F480M is restricted to Visit 5 (taken in parallel to Mask 5) and is shown in red, 
and all other filters with full parallel pointing coverage (F115W, F150W, F200W, F277W, F356W, F444W) are shown in blue. 
} 
\label{fig:parallel_footprints}
\end{figure}

Coordinated parallel NIRCam imaging was taken simultaneously with the primary NIRSpec/PRISM multiobject spectroscopy. 
Altogether imaging was taken in 7 broadband and 2 medium band filters (see Table~\ref{tab:parallel_exposures}), using the MEDIUM8 readout pattern for all exposures. 
This parallel imaging overlaps existing HST/ACS and WFC3 observations (HFF, \citealt{Lotz17}; BUFFALO, \citealt{Steinhardt20}; see Figure~\ref{fig:parallel_footprints}). 
The cumulative exposure time per filter over the parallel footprint ranges from 
0.9 to 5.9 hours, with total areas ranging from 9.2 to 26.9 sq. arcmin. 
This includes imaging in six of the broadband filters that covers the full parallel area (excepting observation issues), and imaging in F090W and the two medium bands F410M and F480M that were only taken in parallel with Mask 5 (see Table~\ref{tab:parallel_exposures}).

\setlength{\tabcolsep}{4pt}
\begin{deluxetable}{ccccc}[t]
\tabletypesize{\footnotesize}
\tablecolumns{5}
\tablewidth{0pt}
\tablecaption{NIRCam Parallel Imaging\label{tab:parallel_exposures}}
\tablehead{
\colhead{Filter} 
& \colhead{Exposure} 
& \colhead{Total Area} 
& \colhead{5$\sigma$ depth}
& \colhead{Masks}
\\
\colhead{(1)} 
& \colhead{(2)}
& \colhead{(3)}
& \colhead{(4)} 
& \colhead{(5)} 
}
\startdata
F090W & 2.8h & 9.2 sq. \arcmin  & 28.94 AB &  5  \\
F115W & 0.9--5.9h  & 26.8 sq. \arcmin  & 28.54 AB  &  1--6 \\ 
F150W & 0.9--5.0h & 25.9 sq. \arcmin  & 28.71 AB &  1--2, 4, 6, 7\tablenotemark{\scriptsize{a}} \\
F200W & 0.9--5.0h & 25.9 sq. \arcmin  & 28.91 AB &  1--2, 4, 6, 7\tablenotemark{\scriptsize{a}} \\
F277W & 0.9--5.9h & 26.9 sq. \arcmin  & 28.96 AB &  1--6\\
F356W & 0.9--5.0h  & 26.2 sq. \arcmin  & 29.02 AB &  1--2, 4, 6, 7\tablenotemark{\scriptsize{a}} \\
F410M & 1.4h & 9.3 sq. \arcmin  & 28.85 AB &  5 \\
F444W & 0.9--5.0h & 26.2 sq. \arcmin & 28.62 AB &  1--2, 4, 6, 7\tablenotemark{\scriptsize{a}} \\
F480M & 1.4h & 9.3 sq. \arcmin  & 28.07 AB &  5 \\
\enddata
\tablecomments{
Depths are calculated within 0.\arcsec16 and 0.\arcsec32 diameter apertures in the short and long wavelength bands, respectively, using noise properties derived from the weight maps and corrected to total assuming a point source geometry. 
As the footprint is inhomogenous, these estimates correspond to a 0.7 arcmin$^2$ box centered at $(3.6012969,-30.4908199)$. 
\\
Columns: 
(1) NIRCam filter.  
(2) Filter exposure time across footprint (hours). 
(3) Total filter footprint area (sq. arcmin).
(4) Imaging $5\sigma$ depth. 
(5) Mask(s) with which the filter was observed in parallel.  
}
\tablenotemark{\scriptsize{a}}:  
Parallel imaging in F150W, F200W, F356W, F444W in Visit 3 (Mask 3) was lost due to a SSR drive exception (see Sec.~\ref{sec:design}). 
Repeat observations were taken on 31 July 2024. 
\end{deluxetable}
\setlength{\tabcolsep}{2pt}

\section{Updates to the UNCOVER strong lensing model of Abell 2744} 
\label{app:lens-model}
We use the UNCOVER spectroscopy, described in this work, as well new JWST/NIRCam imaging \citep[][]{Suess24} and grism spectroscopy (R. Naidu \& J. Matthee, et al., in prep.) of the Abell 2744 field to update the UNCOVER strong lensing (SL) model of the cluster, as presented in Section~\ref{app:SL-model}. 

The parametric lens model of Abell~2744 is constructed with an updated version of the \citet{Zitrin15} analytical method. It comprises five smooth cluster-scale dark matter halos centered on each of the sub-clusters' BCG, modeled as pseudo-isothermal elliptical mass distributions \citep[PIEMDs;][]{Kassiola93}, and 552 cluster member galaxies (see Appendix~\ref{app:cluster-members}), modeled as dual pseudo-isothermal ellipsoids \citep[dPIEs;][]{Eliasdottir07}. We refer the reader to \citet{Furtak23} for more details on the implementation and setup of our Abell~2744 model.

While the currently available \texttt{v1.1} SL model presented in \citet{Furtak23} is based on HST-selected cluster members and mostly photometric multiple image systems in the northern and north-western extended cluster sub-structures, the \texttt{v2.0} model presented here adds additional cluster member galaxies selected with JWST (Appendix~\ref{app:cluster-members}) and new spectroscopic redshifts of multiple image systems as constraints (section~\ref{app:multiple-images}). The new SL model (Section~\ref{app:SL-model}) maps are also made public on the UNCOVER website in the framework of DR4 (\url{https://jwst-uncover.github.io/DR4.html}; see Section~\ref{app:SL-model}).

\subsection{JWST cluster member selection}
\label{app:cluster-members}

\begin{figure}
    \centering
    \includegraphics[width=\columnwidth]{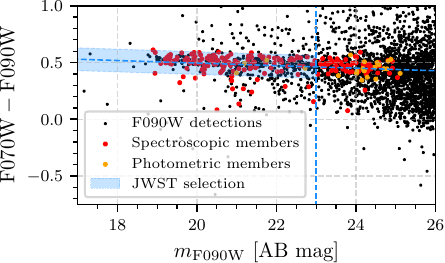}
    \caption{JWST/NIRCam color-magnitude diagram of objects detected in Abell 2744, showing the cluster's red sequence. Known spectroscopic and photometric cluster members from \citet{Bergamini23a} are shown as red and orange dots and our red sequence selection is shown as the blue shaded area.}
    \label{fig:MegaScience_cm-diagram}
    \vspace{10pt}
\end{figure}

Thanks to the JWST \textit{Medium Bands, Mega Science} program \citep[\textit{MegaScience};][]{Suess24}, we now have NIRCam F070W and F090W imaging data covering the entire UNCOVER field at our disposal. These two filters straddle the 4000\,\AA\ break at the cluster's redshift $z_{\mathrm{d}}=0.308$ and are therefore ideally suited for photometric selection of cluster members from the red sequence \citep[e.g.][]{Repp18}. We use \texttt{SExtractor} \citep{Bertin96} in dual-imaging mode to detect sources in the F090W mosaic and measure their photometry in F070W and F090W. Following our approach in \citet{Furtak23} and \citet{Furtak24c}, we then use the colors of the known spectroscopic cluster members from \citet{Bergamini23a} to calibrate the cluster's red sequence in the color-magnitude diagram (see Figure~\ref{fig:MegaScience_cm-diagram}). Cluster members are then selected in a color-window of width 0.1 around the red sequence and brighter than 23\,magnitudes in the F090W band. The resulting sample is cross-matched with the known spectroscopic and HST-selected cluster members \citep[][]{Furtak23} to make sure no galaxy is counted doubly.

As a result, we complement our previous cluster member sample from \citet{Furtak23} with 132 new NIRCam selected cluster members. This bring the total number of cluster members included in the SL model to 552, now spanning the entire 45\,arcmin$^2$ of the UNCOVER field. The new, NIRCam-selected sample in particular adds cluster members in the north-east to north-west of the cluster, areas which were not covered with HST.

\begin{deluxetable}{cccr}
\label{tab:multiple-images}
\tablecolumns{4}
\tablecaption{New spectroscopic redshifts of multiply-imaged sources included in our \texttt{v2.0} SL model of Abell 2744.}
\tablehead{
\colhead{System ID} 
& \colhead{MSA ID} 
& \colhead{$z_{\mathrm{spec}}$}
& \colhead{Redshift reference}\\
\colhead{(1)} 
& \colhead{(2)}
& \colhead{(3)} 
& \colhead{(4)}
}
\startdata
    \multicolumn{4}{c}{\textit{UNCOVER spectroscopy}}\\\hline
    53  &   13123   &   7.045   &   \citet{Furtak23b}.\\
    65  &   60046   &   3.519   &   This work.\\
    67  &   33295   &   2.322   &   Siegel et al. (in prep.).\\
    69  &   29315   &   2.411   &   This work.\\
    70  &   60053   &   2.392   &   This work.\\
    72  &   60061   &   3.747   &   This work.\\
    74  &   60067   &   2.374   &   This work.\\
    78  &   60018   &   2.315   &   This work.\\
    80  &   60010   &   3.672   &   Williams et al. (in prep.).\\
    81  &   60081   &   3.479   &   This work.\\
    86  &   16155   &   6.875   &   \citet{Atek24}.\\\hline
    \multicolumn{4}{c}{\textit{ALT spectroscopy}}\\\hline
    84  &   11254   &   6.873   &   R. Naidu \& J. Matthee, et al. (in prep.).\\
    85  &   -       &   4.753   &   R. Naidu \& J. Matthee, et al. (in prep.).\\\hline
    \multicolumn{4}{c}{\textit{VLT/MUSE spectroscopy}}\\\hline
    68  &   -       &   2.584   &   \citet{Bergamini23b}.\\
\enddata
\tablecomments{A full table of multiple images used in the \texttt{v2.0} model is included in the public SL model release at \url{https://jwst-uncover.github.io/DR4.html}.\\
Columns: 
(1) ID number of the multiple image system. 
(2) ID number of the MSA-slit on one of the images.
(3) Spectroscopic redshift. 
(4) Reference to the spectroscopic redshift measurement.}
\end{deluxetable}

\begin{figure*}
    \centering
    \includegraphics[width=\textwidth]{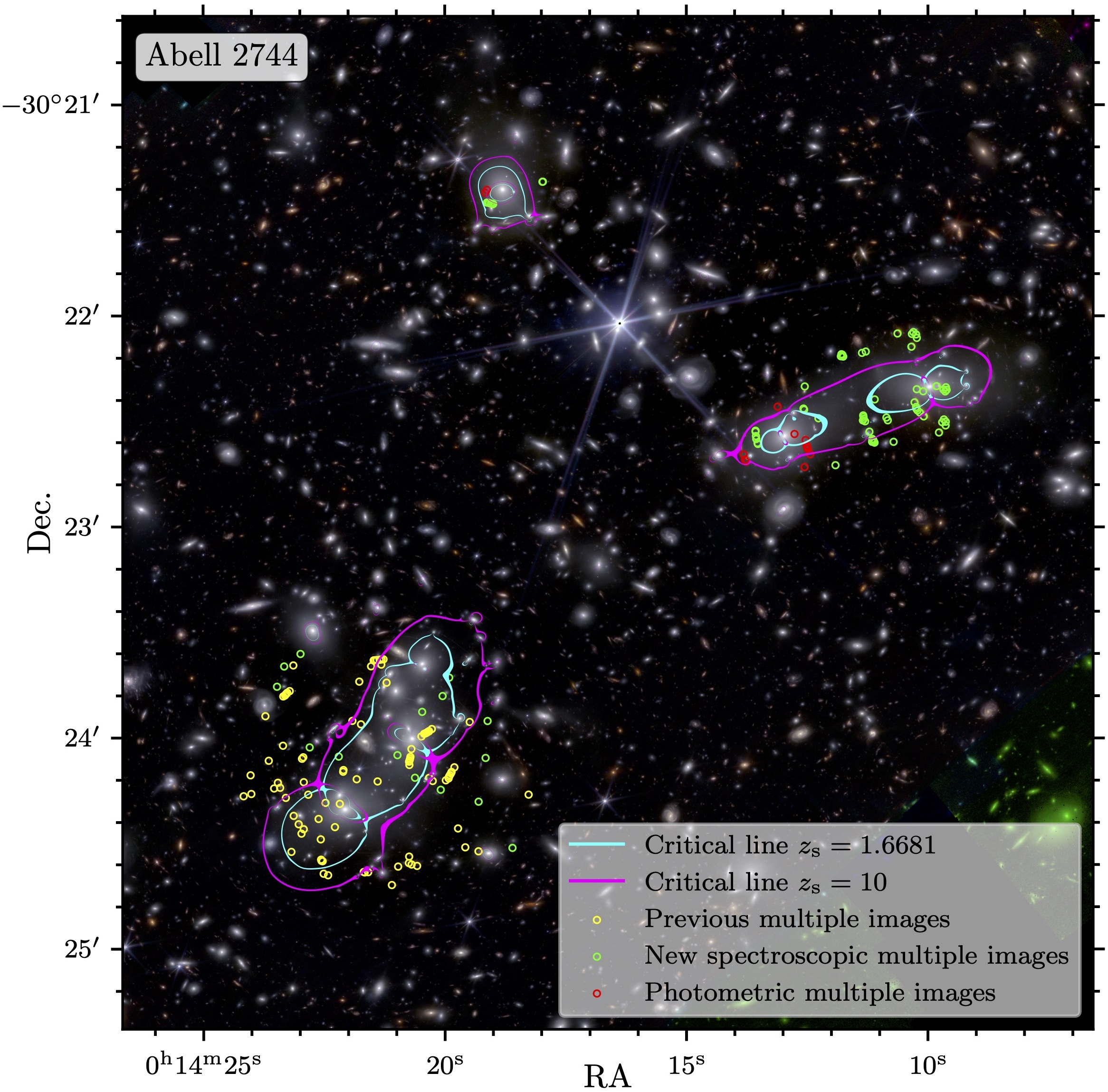}
    \caption{A $4.3\arcmin\times4.8\arcmin$ cutout of an UNCOVER and \textit{MegaScience} NIRCam composite-color image of Abell 2744 including all broad and medium bands. Overlaid we show the critical curves of our SL model for source redshifts $z_{\mathrm{s}}=1.6881$ (corresponding to system~1) and $z_{\mathrm{s}}=10$ in blue and purple respectively. Multiple images from \citet{Bergamini23a}, used with spectroscopic redshifts in our \texttt{v1.0} model, are shown in yellow and multiple images with new spectroscopic redshifts in our \texttt{v2.0} model are shown in green. Photometric multiple images are shown in red. The area between the main cluster and the north-western sub-structure in particular has high magnifications of order $\mu\gtrsim4$ for sources at $z_{\mathrm{s}}=10$ (see Fig.~\ref{fig:magnification-map}). Note, a vectorized full 0.04\arcsec/pix resolution version of this figure is included in the public \texttt{v2.0} SL model release.}
    \label{fig:crit-lines}
\end{figure*}

\subsection{New spectroscopic redshifts of multiple images} 
\label{app:multiple-images}

The unprecedented depth and areal coverage of the UNCOVER survey's imaging \citep[][]{Bezanson24} enabled us to detect new multiple image systems in north-western and northern extensions of Abell 2744 which were previously not know to be dense enough to produce strong lensing \citep[][]{Furtak23}. These new systems were however not constrained with spectroscopic redshifts in the first UNCOVER model (\texttt{v1.0}) due to lack of spectroscopic coverage in those areas. Multiple images without precise redshift information are known to significantly bias SL models of galaxy clusters \citep[e.g.][]{Johnson16} which is why spectroscopic redshifts are paramount for accurate SL modeling and magnification estimates.

After the publication of our \texttt{v1.0} model \citep{Furtak23}, new VLT/MUSE observations found system 68 to lie at $z_{\mathrm{spec}}=2.584$ \citep{Bergamini23b}. We included that new redshift in our \texttt{v1.1} model release in June~2023, but the model remained mostly constrained with photometric systems in the north-west and the north. With the UNCOVER JWST/NIRSpec observations presented in this work, we are now able to spectroscopically confirm numerous multiple image systems in the whole UNCOVER field. In total, we obtained 10 new spectroscopic redshifts. These in particular include the triply-imaged high-redshift AGN A2744-QSO1 at $z_{\mathrm{spec}}=7.045$ \citep[system 53;][]{Furtak23b}, a low-mass heavily star-forming object at $z_{\mathrm{spec}}=6.875$ \citep[system 86;][]{Atek24}, and a massive quiescent galaxy at $z_{\mathrm{spec}}=2.322$ stretched into an arc (system 67; Siegel et al. in prep.). In addition, the JWST Cycle~2 program \textit{All the Little Things} (ALT; Program-IS: 3516 PIs J.~Matthee \& R.~Naidu) observed the Abell 2744 field with JWST/NIRCam grism spectroscopy in the F356W filter, which enabled the discovery of two new multiple image systems at $z_{\mathrm{spec}}=6.873$ (system 84) and $z_{\mathrm{spec}}=4.753$ (system 85) respectively (R. Naidu \& J. Matthee, et al., in prep.), which we also included in the model. Note that system 84 also has an UNCOVER NIRSpec redshift which agrees with the ALT redshifts.

We list all new multiple image redshifts in Table~\ref{tab:multiple-images} and show them in Figure~\ref{fig:crit-lines}. In total, our new \texttt{v2.0} SL model is constrained by 187 multiple images belonging to 66 individual sources. Of these, 60 sources now have spectroscopic redshifts, leaving only 6 multiply-imaged sources with free redshifts in the model. For additional constraining power, we now also use parity information of 4 very close knot systems, systems 65.3, 67.3, 78.3, and 80.2, as constraints in the model \citep[see equations~8 and~9 in][]{Furtak23}. A full list of multiple images, including coordinates and redshifts, is included in the public \texttt{v2.0} SL model release.

\bibliography{refs, refs_software}{}
\bibliographystyle{aasjournal}

\end{document}